\documentclass[aps,prl,floatfix,twocolumn,showpacs,superscriptaddress]{revtex4}

\usepackage{graphicx}
\usepackage{amssymb}
\usepackage{amsmath}

\newcommand {\ket}[1] {|#1 \rangle}
\newcommand {\bra}[1] {\langle#1 |}

\begin{document}

\title{Quantum computing with alkaline earth atoms}

\author{Andrew J.~Daley}
\affiliation{California Institute of Technology, Pasadena, California 91125, USA}
\affiliation{Institute
for Theoretical Physics, University of Innsbruck, A-6020 Innsbruck, Austria\\ and Institute for Quantum Optics and Quantum Information of the
Austrian Academy of Sciences, A-6020 Innsbruck, Austria} 

\author{Martin M. Boyd}
\affiliation{JILA, National Institute of Standards and Technology and University of Colorado,  Boulder, CO 80309-0440, USA}

\author{Jun Ye}
\affiliation{California Institute of Technology, Pasadena, California 91125, USA}
\affiliation{JILA, National Institute of Standards and Technology and University of Colorado,  Boulder, CO 80309-0440, USA}

\author{Peter Zoller}
\affiliation{California Institute of Technology, Pasadena, California 91125, USA}
\affiliation{Institute
for Theoretical Physics, University of Innsbruck, A-6020 Innsbruck, Austria\\ and Institute for Quantum Optics and Quantum Information of the
Austrian Academy of Sciences, A-6020 Innsbruck, Austria} 

\date{August 14, 2008}

\begin{abstract}
We present a complete scheme for quantum information processing using the unique features of alkaline earth atoms. We show how two completely independent lattices can be formed for the $^1$S$_0$ and $^3$P$_0$ states, with one used as a storage lattice for qubits encoded on the nuclear spin, and the other as a transport lattice to move qubits and perform gate operations. We discuss how the $^3$P$_2$ level can be used for addressing of individual qubits, and how collisional losses from metastable states can be used to perform gates via a lossy blockade mechanism.
\end{abstract}


\maketitle

First steps in implementing quantum information processing with neutral atoms have been taken in experiments with alkali atoms. These have demonstrated basic building blocks including entangling gates with coherent collisions in optical lattices \cite{exchangegate,spindep1}, Rydberg states \cite{rydberg}, and cavity quantum electrodynamics \cite{cqed,quantumnetwork}, as well as high-fidelity register loading \cite{loading, esslinger04}. 
Challenges in the further development of neutral atom systems towards scalable quantum computing include single qubit addressing, and the achievement of high fidelity operations whilst avoiding decoherence, e.g., due to magnetic field fluctuations \cite{spindep1}. Alkaline earth(-like) atoms, as developed in the context of optical clocks \cite{opticalclock}, and degenerate gases \cite{takasu}, offer unique and novel opportunities to address these challenges \cite{aeqc1,aeqc2}. The advantages include the possibility to encode qubits in nuclear spin states, decoupled from the electronic state in both the $^1$S$_0$ ground state and the very long lived $^3$P$_0$ metastable state on the clock transition \cite{aeqc1}. 
We show below that these ground and excited states can be manipulated completely independently by laser light, allowing the construction of independent optical lattices for the two states. This leads to a quantum computing scenario where qubits are stored in long lived states in a \textit{storage} lattice (associated with the $^1$S$_0$ ground state), and can be transferred with individual addressing to a \textit{transport} lattice (associated with the $^3$P$_0$ metastable state). This can be used to move qubits around, and perform high-fidelity entangling gate operations (see Fig.~1a), or also many such operations in parallel \cite{spindep2,spindep1}. We discuss a complete quantum computing proposal in this context, with quantitative analysis for $^{87}$Sr \cite{opticalclock}. This toolbox of techniques developed here is also of immediate relevance for quantum simulation.

\begin{figure}[tb]
\includegraphics[width=8.5cm]{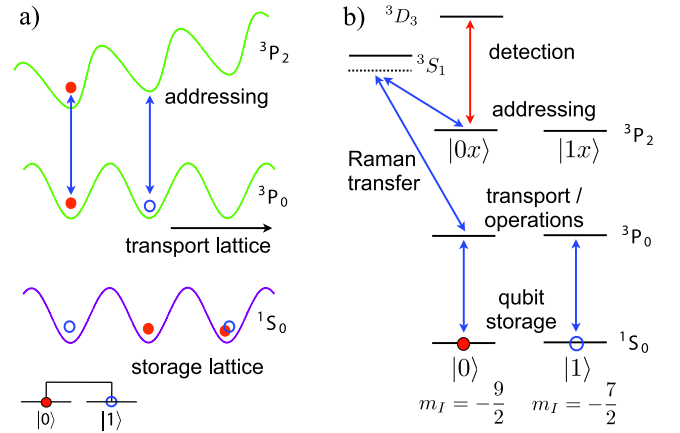}
\caption{Quantum computing with independent lattices: (a) Qubits in long-lived states in a storage lattice are transferred to a completely independent transport lattice for gate operations between distant qubits, or addressed individually by coupling to a level that is shifted by a gradient field. (b) This can be accomplished by encoding qubits in nuclear spin states, producing independent lattices for the $^1$S$_0$ and $^3$P$_0$ levels, and using $^3$P$_2$ for individual addressing.} \label{fig1}
\end{figure}

The details of our scheme are shown schematically in Fig.~1. In a large magnetic field, the nuclear spin decouples from the electronic state on the clock transition $^1$S$_0$--$^3$P$_0$. We can then encode qubits in nuclear spin states of different magnetic quantum number $m_I$ (e.g., for $^{87}$Sr, we can define $\ket{0}\equiv\ket{^1{\rm S}_0, m_I=-9/2}$, and $\ket{1}\equiv\ket{^1{\rm S}_0, m_I=-7/2}$, see Fig.~1b). These states are very insensitive to magnetic field fluctuations. Because the $^1$S$_0$ ground state and $^3$P$_0$ metastable state (with lifetime $\sim 150$s for $^{87}$Sr) belong to different transition families and are separated by optical frequencies, we can search for two wavelengths where an optical field will generate an AC-Stark shift for each of these states completely independently of the other, as shown in Fig.~2a. In Fig.~2b,c we plot the polarisability of the $^1$S$_0$ and $^3$P$_0$ states of $^{87}$Sr at different wavelengths computed from oscillator strengths in Ref. \cite{porsev}. We see very clearly that at 627nm, the polarisability of $^3$P$_0$ is zero because of cancelling shifts of different signs from more highly excited triplet levels, whilst the polarisability of $^1$S$_0$ is $\sim 430$a.u.. Thus, we can form a deep optical lattice  (where tunnelling is negligible on experimental timescales) at a wavelength of 627nm as a \textit{storage} lattice for qubits, which will not affect the $^3$P$_0$ states. Similarly, the polarizability of $^3$P$_0$ at 689.2nm is $\sim 1550$a.u, whereas the polarisability of $^1$S$_0$ is zero. This is largely because of the near-resonant coupling of $^1$S$_0$ to $^3$P$_1$, which is made possible without large spontaneous emission rates due to the narrow linewidth of $^3$P$_1$. This lattice can be used for \textit{transport}, and atoms in it will not be affected by the storage lattice. These lattices can be made to have the same spatial period by using angled beams in the case of the 627nm light, so that the lattice period is increased to match that formed by counterpropagating beams at 689.2nm, and the depths can be made equal by using light of intensity $I_0$ for the storage and $\sim I_0/4$ for the transport lattice, facilitating transfer of atoms between the two lattices. Gate operations can then be performed between distant sites by transfering atoms state-selectively into the transport lattice, and moving them to the appropriate distant site (see below for more details). This is somewhat reminiscent of the use of spin-dependent lattices for alkali atoms \cite{spindep1,spindep2}, where lattice lasers are tuned between fine-structure states, which can lead to large heating and decoherence from spontaneous emissions. Here, the lattices can be made completely independent by selection of the correct wavelengths.

\begin{figure}[tb]
\includegraphics[width=8.5cm]{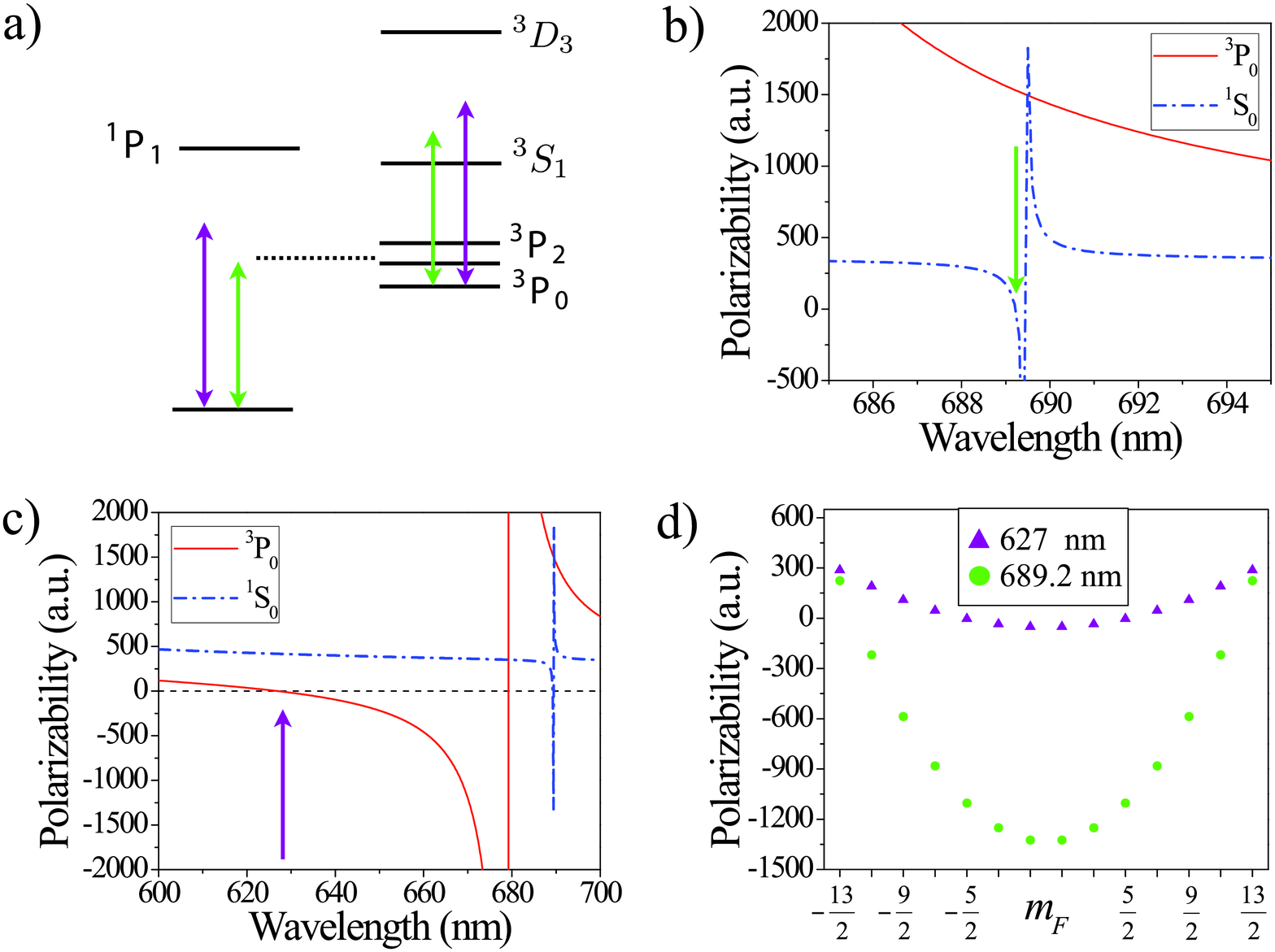}
\caption{(a) Independent optical lattices can be produced for the $^1$S$_0$ and $^3$P$_0$ levels by finding wavelengths where the polarisability of each of the levels is zero and the other non-zero. (b) AC polarisability of $^1$S$_0$ and $^3$P$_0$ levels near $689$nm.  (c) AC Polarisability of $^1$S$_0$ and $^3$P$_0$ levels near $627$nm. (d) AC polarisability of different $m_F$ sublevels of the $^3$P$_2$, $F=13/2$ hyperfine level for $\pi-$polarised light at 627nm and 689.2nm.} \label{fig2}
\end{figure}

An essential ingredient for general-purpose quantum information processing is the \textit{individual addressing} of qubits, both for readout and gate operations, which can be achieved in this system by coupling selectively to states in the long-lived $^3$P$_2$ manifold. As shown schematically in Fig.~1, we would transfer qubit states $\ket{0}$ and $\ket{1}$ to the $^3$P$_0$ level (which can be done state-selectively in a large magnetic field due to the differential Zeeman shift of $109$Hz/G between $^1$S$_0$ and $^3$P$_0$), and then selectively transfer them to additional \textit{readout} levels $\ket{0x}$ and $\ket{1x}$ in the $^3$P$_2$ level (e.g., for $^{87}$Sr we could choose $\ket{0x}\equiv\ket{^3{\rm P}_2,F=13/2,m_F=-13/2}$ and $\ket{1x}\equiv\ket{^3{\rm P}_2,F=13/2,m_F=-11/2}$, where $F$ is the total angular momentum and $m_F$ the magnetic sublevel, and connect these states to the $^3$P$_0$ level via off-resonant Raman coupling to a $^3$S$_1$ level). 
The individual qubit selectivity can be based on a gradient magnetic field, as $^3$P$_2$ is much more sensitive to magnetic fields $^3$P$_0$ or $^1$S$_0$. A gradient field of 100 G/cm will provide an energy gradient of 410 MHz/cm for $\ket{0x}$ or an energy difference of about 15kHz between atoms in neighbouring sites. In the same field the $^3$P$_0$ level states will be shifted by $-m_I \times 1$Hz in neighbouring sites. This not only enables selectivity, but means that additional relative phases collected between qubit states in the storage or transport lattices will be small on the timescale of the transfer operation \footnote{If necessary, these phases can be made homogeneous by applying the gradient shift twice, the second time with the opposite gradient.}, which again indicates the advantage of storing qubits on the nuclear spin states. This addressing can be used to perform qubit readout, by selectively transferring only the $\ket{0}$ state, then making fluorescence measurements of the occupation of the $^3$P$_2$ level (e.g., using the cycling transition between the $^3{\rm P}_2, F=13/2$ and $^3{\rm D}_3$ manifold). It can also be used to transfer atoms site-dependently to the transport lattice, by first selectively transferring atoms to the $^3$P$_2$ level, then returning the remaining atoms to $^1$S$_0$, before coupling the atoms in $^3$P$_2$ back to $^3$P$_0$.

A necessary requirement here is that our states $\ket{0x}$ and $\ket{1x}$ are trapped in the combination of the storage and transport lattices (these will both provide AC-Stark shifts for the $^3$P$_2$ level). In Fig.~2d we plot the polarisability of all of the magnetic sublevels of $^3$P$_2$, $F=13/2$ at our lattice wavelengths, and the large tensor shifts make certain $m_F$ levels suitable for trapping at the same locations as our qubit states. If the depths of the storage and transport lattices are chosen to be equal, then the intensity of the lattice at 689.2nm will be a factor of four smaller than that at 627nm, and both the $\ket{0x}$ level and the $\ket{1x}$ will be trapped, in lattices about 2/3 and 1/3 of the depth of the storage lattice respectively. In all cases, the timescale for transfer processes $\tau_{\rm transfer}$ is limited by the smallest trapping frequency $\omega_t$ out of the two potentials to and from which the atom is being transferred (so that atoms are not coupled to excited motional states), and by the frequency shift $\omega_e$ between neighbouring sites in the case of position-selective transfer, as $\tau_{\rm transfer} \gg \max(2\pi /\omega_t,2\pi/\omega_e$). Note that an alternative to magnetic field gradients for addressing would involve applying a laser with spatially varying intensity at the so-called magic wavelength (for equal shifts of the $^3$P$_0$ and $^1$S$_0$ levels), which would provide position-dependent differential AC-Stark shifts between $^3$P$_0$ and $^3$P$_2$.

\begin{figure}[tb]
\includegraphics[width=8.5cm]{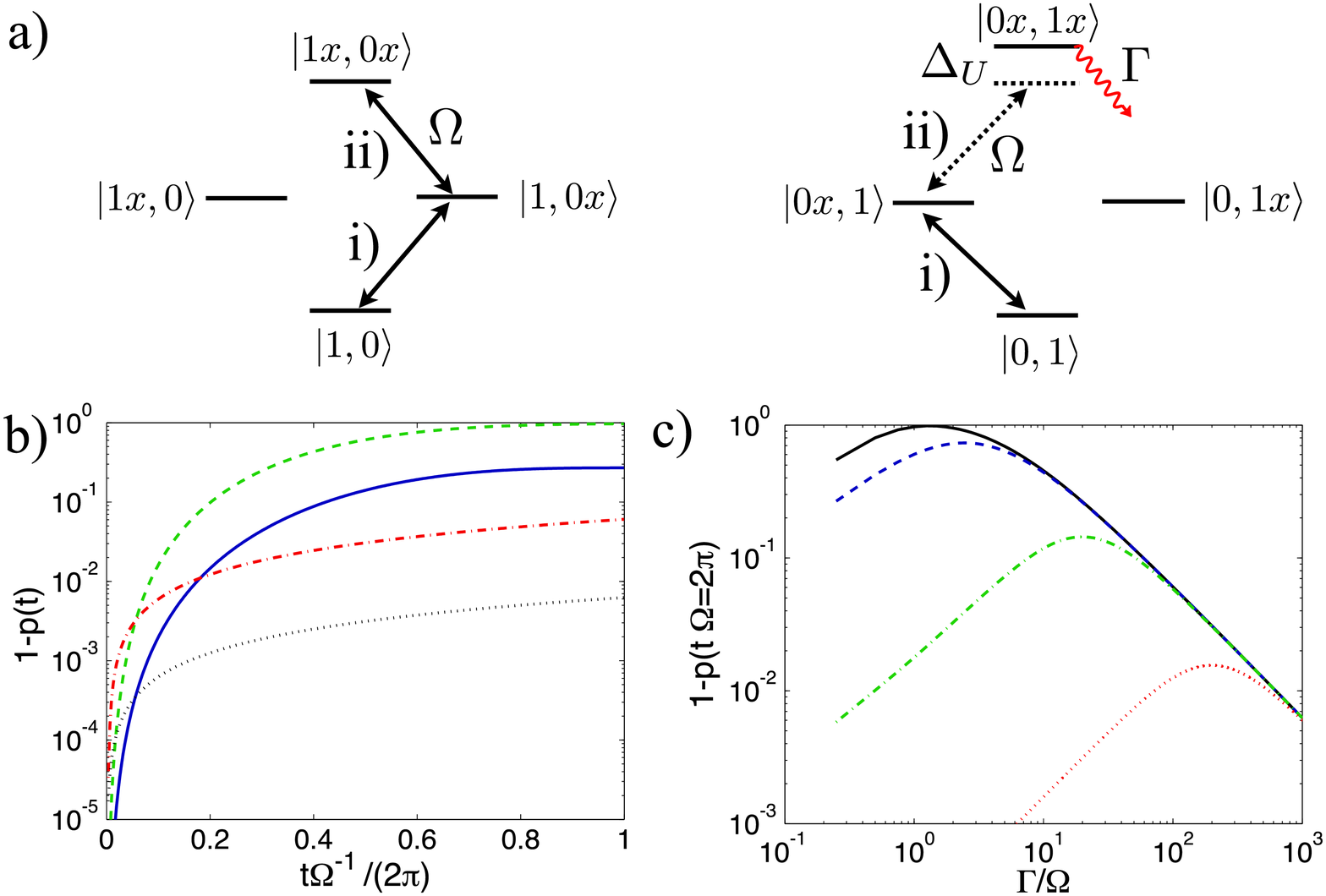}
\caption{(a) Two-qubits levels in a lossy blockade gate, contrasting behaviour for the initial state $\ket{1,0}$, where the atoms are separated, and $\ket{0,1}$, where the atoms undergo collisions in the excited state on the same site. Atoms are (i) excited from the state $\ket{0}\rightarrow\ket{0x}$, and then (ii) coupled from $\ket{1}\rightarrow\ket{1x}$. The second process is blocked for the initial state $\ket{0,1}$ by elastic and inelastic collisional interactions. (b) Loss probability during step (ii) up to time $t$ with the initial state $\ket{0x,1}$ for $\Gamma/\Omega=$1 (solid line), 10 (dashed), 100 (dash-dot), and 1000 (dotted), with $\Delta_U=0$. (b) Loss probability up to the gate completion time $\Omega t =2\pi$ for $\Delta/\Omega=$0 (solid), 1 (dashed), 10(dash-dot), and 100(dotted).
}  \label{fig3}
\end{figure}

Single-qubit gates can be performed either directly via a Raman coupling of the qubit states after transferring them to the $^3$P$_0$ level, or with single-qubit addressability by making use of the $^3$P$_2$ level. 
Two-qubit gates are then performed using the transfer lattice. In particular, a phase gate between qubits in site $i$ and $j$ can be performed in a straight-forward manner by: (i) transferring atoms in $\ket{0}$ on site $i$ (and $j$) to the transport lattice; (ii) moving the transport lattice relative to the storage lattice so that an atom that was originally in the $\ket{0}$ state on site $i$ would now be present at site $j$; (iii) generating a phase $\phi$ for the state conditioned on whether two atoms are on the same site \footnote{Here it may be useful to transfer the $\ket{0}$ qubit state from site $j$ to an unoccupied site.}; and (iv) returning the atoms to their original position. In this protocol, if we express the state of the qubits in sites $i$ and $j$ in the basis $\ket{q_i ,q_j}$, then the state $\ket{0,1}\rightarrow \exp(i\phi) \ket{0,1}$, and all other states are unchanged. Many such phase gates can also be performed in parallel \cite{spindep2}, For example, cluster states \cite{cluster} could be produced in a single operation entangling all atoms in neighbouring sites.

The phase in step (iii) can be generated by an onsite collisional shift $U$ if the scattering length between two atoms in any combination of the $^1$S$_0$ and $^3$P$_0$ levels is significant. Then $\phi=UT$, where the atoms are left onsite for a time $T$, analogously to previous gates by controlled collisions in alkali atoms \cite{spindep1,spindep2}. For Sr atoms the collisional interactions are normally weak, but could be increased using optical Feshbach resonances \cite{srfeshbach}. However, this also motivates the consideration of other gate schemes, especially those based on interactions in the $^3$P$_2$ manifold. We will discuss here blockade mechanisms with both coherent interactions and lossy channels. For sufficiently large onsite magnetic dipole-dipole interactions, which provide an energy shift $\Delta_U$ between $^3$P$_2$-$^3$P$_2$ and $^3$P$_0$-$^3$P$_2$ onsite collisional interactions, we can use a dipole blockade mechanism to produce a $\phi=\pi$ phase shift, as proposed, e.g., for Rydberg atoms \cite{int:rydberg}: (i) excite all $\ket{0}$ qubit states to an auxillary level $\ket{0x}$ with a $\pi$-pulse (ii) couple all $\ket{1}$ qubit states to an auxillary level $\ket{1x}$ with a $2\pi$-pulse at Rabi frequency $\Omega$, assuming that there is no collisional interaction between the $\ket{0x}$ state and either $\ket{1}$ or $\ket{1x}$. If the two atoms are on the same site the coupling is detuned by a frequency $\Delta_U$ and the transfer is blocked; (iv)  Return the $\ket{0x}$ state to the $\ket{0}$ state with a $\pi$ pulse. This is shown schematically in Fig.~3a for the states $\ket{01}$ and $\ket{10}$.  For two atoms not on the same site, two phases of $\pi$ are collected, so that $\ket{q_i ,q_j}\rightarrow \ket{q_i,q_j}$. However, the state $\ket{0,1}$ collects only a phase $\phi=\pi$ due to the Rabi flop of only one atom, as the second transfer is blocked. In practice, the state $\ket{0,1}$ will collect a small additional phase $\phi\sim \Omega/\Delta_U$.

In addition to large elastic interactions, we expect large inelastic spin-flip losses in the $^3$P$_2$ manifold, as discussed in Refs. \cite{greene}, which could reduce the fidelity of the blockade gate. However, this loss can actually help us in producing the blockade effect, as large losses at a rate $\Gamma$ from a given level can also dynamically suppress occupation of that level, based on a similar mechanism to that observed recently in experiments with cold molecules \cite{durrrempe}. In the limit $\Delta_U \ll \Gamma$, one could even produce a gate based entirely on a lossy blockade mechanism. This is made possible because the energy change in the inelastic collision is larger than the lattice depth, so that we can assume that atoms are untrapped by the lattice, and coupled to a continuum of motional states. We can estimate the loss rates in the lattice based on the free-space values \cite{greene}, assuming that the length scale on which the physics of the inelastic collision takes place is smaller than the confinement length in a lattice site, and that the collisions are thus unaffected by the presence of the lattice. The on-site loss rates from $^3$P$_2$ could then reach values of the order of $\Gamma=2\pi\times20$kHz for lattice densities of $10^{16}$cm$^{-3}$.

In the presence of loss, the basic physics of the second step of the protocol, as illustrated in Fig.~3a then reduces to a two level system, coupling the states $\ket{g}\equiv\ket{0x,1}$ and $\ket{e}\equiv\ket{0x,1x}$. The non-Hermitian effective Hamiltonian describing the loss process is given by 
\begin{equation}
H=\frac{\Omega}{2} (\ket{e}\bra{g} + \ket{g}\bra{e}) +(- \Delta_U -i \frac \Gamma 2) \ket{e}\bra{e}.
\end{equation}
In the limit $|\Delta_U+i \Gamma/2| \gg \Omega$, we can describe the time evolution of a system initially prepared in $\ket{g}$ in second order perturbation theory, giving the probability that no decay has occurred at short times $t$ as
$p={\rm e}^{-\Gamma_{\rm eff}t}$,
with
$\Gamma_{\rm eff}\approx \Omega^2\Gamma/[4(\Delta_U^2+\Gamma^2/4)]$. For our lossy blockade gate the largest probability of loss occurs in the regime $\Gamma \gg \Delta_U$, where the ratio of the loss time to the gate time (determined by $\Omega$) is given by
$\tau_{\rm loss}/\tau_{\rm gate}=\Omega/\Gamma$. This will limit the fidelity of the lossy blockade gate to $1-\Omega/\Gamma$, provided that there are no additional collisional shifts. If $\Delta_U\neq 0$, then the loss probability during the gate is decreased, as shown in Fig.~3b, and the gate fidelity is correspondingly higher.

The fidelity of our gates and storage lifetime of our qubits is high due to the encoding of qubits in the nuclear spin states. For magnetic field fluctuations $\Delta B < 10^{-3}$G, the corresponding differential shift of the qubit states is $\Delta\omega_{B}<0.3$Hz, as the Zeeman shift is $-185$Hz/G in the $^1$S$_0$ level, and $-295$Hz/G in the $^3$P$_0$ level. This is suppressed by over three orders of magnitude compared with electron spin states. Relative intensity fluctuations in the storage and transport lattices will cause changes in the ground state energy of states in different lattices, but if this is controlled to one part in $10^6$, the relative shifts $\Delta \omega_{\rm intensity}<0.05$Hz. In the presence of both the storage and trapping lattices, each with a trapping frequency of $25$kHz, the spontaneous emission lifetimes of the various levels are: $^1$S$_0:\sim 20$s, $^3$P$_0:\sim 2$s, $^3$P$_2:\sim 1$s. These constitute the largest source of decoherence during gate operations, but the associated timescales are much larger than the gate times, which in the worst case are limited by the trap frequency to be a few ms. We expect, therefore, that gate fidelities $\mathcal{F}> 99\%$ can be achieved in experiments. Similarly, collisional losses from metastable states, which occur only when two atoms are brought onto a single site, should play a small role except during lossy blockade gates, as discussed above. The collisional loss rates from $^3$P$_0$ levels, which could play a role during the blockade gate operation are not yet known, however for gate times on the order of $1$ms, we require collisional stability of our atoms only for timescales of $100$ms in order to achieve gate fidelities $\mathcal{F}>99$\%.  On the other hand, if losses from the $^3$P$_0$ are large, then these could also be directly used for a lossy blockade gate with the two atoms being coupled from $^1$S$_0$ to $^3$P$_0$. 
  
As the isotopes of Sr or Yb with non-zero nuclear spin are fermionic, we have a substantial advantage in loading a quantum register with one atom per lattice site. If the lattice is ramped up adiabatically in the presence of a degenerate Fermi gas, a band insulator will form \cite{esslinger04} provided that the temperature is smaller than the lattice bandgap, and sites with missing atoms will typically be localised near the edges of any external trapping potential \cite{calarco04}, leaving a regular array in the centre of the trap. Moreover, because we have two internal states trapped by independent lattices, this system would be an ideal candidate for improvment of the quantum register by coherent filtering \cite{rablloading} or implementation of a fault-tolerant dissipative loading scheme \cite{agloading}. The latter involves transferring atoms from a reservoir where the atoms are in an internal state that is not trapped by the lattice into a state where they are trapped by the lattice, which could easily be achieved here by initially using a lattice that traps only one internal state. 

This is a complete quantum computing proposal making use of the unique features of alkaline earth atoms. In addition, the optical clock transition and nuclear spin states provide a
natural basis for interfacing stationary (nuclear) and flying (photonic) qubits \cite{quantumnetwork}. The clean realisation of state-dependent lattices also opens a toolbox of techniques for quantum simulation \cite{opticallattices}, with such applications as implementation of spin models in optical lattices \cite{spindepapp}, or investigating dissipative dynamics with a reservoir gas coupled to atoms in an optical lattice \cite{reservoir}.

\begin{acknowledgements}
We thank C. Greene, M. Lukin, and A. Gorshov for useful discussions. AJD thanks the Institute for Quantum Information at Caltech for support, and AJD, JY, and PZ thank Caltech for hospitality. Work at JILA is supported by DARPA, NIST, and NSF, and work in Innsbruck is supported by the Austrian Science Foundation (FWF), and by the EU Network NAMEQUAM. 
\end{acknowledgements}

\end{document}